\documentclass{appolb}
\usepackage{amssymb,amsmath}
\usepackage{hyperref}
\usepackage[numbers]{natbib}
\usepackage{graphbox}
\usepackage{grffile}

\usepackage{graphicx}
\usepackage{xspace}
\newcommand{\shine}{NA61/SHINE\xspace}
\newcommand{\ts}{\textsuperscript}

\begin{document}
\title{NA61/SHINE results on search for critical point%
\thanks{Presented at XXXII Cracow Epiphany Conference on the recent results from Heavy Ion Physics, 12-16 Jan 2026, IFJ PAN, Krak\'ow, Poland}%
}
\author{Nikolaos Davis
\address{The Henryk Niewodnicza\'nski Institute of Nuclear Physics PAN,\\ ul. Radzikowskiego 152\\
31-342 Krak\'ow, Poland}
\\[3mm]
for the \textsc{NA61/Shine} Collaboration
}
\maketitle
\begin{abstract}
The \shine experiment at the CERN SPS is a multipurpose fixed-target spectrometer for charged and neutral hadron measurements. Its research program includes studies of strong interactions as well as reference measurements for neutrino and cosmic-ray physics. One major goal of its strong interaction program is to determine the existence and pinpoint the location of the QCD critical point, an object of both experimental and theoretical studies.

This contribution will summarize the current status of \shine critical point searches in nucleus-nucleus collisions, in the collision energy range $\sqrt{s_{NN}} = 5-17$~GeV. The review includes studies of fluctuations of net-electric charge, femtoscopy analysis of $\pi-\pi$ pairs, as well as intermittency of protons and negatively charged hadrons. No clear indication of the critical point has been observed so far. Finally, we report on the development of novel methods aimed at solving the long-standing problem of bin-by-bin correlations in experimental intermittency analysis, and for a more accurate handling of systematics and uncertainties.
\end{abstract}
  
\section{Introduction}

\begin{figure}
 \begin{center}
    \includegraphics[align=c,height=0.18\textheight]{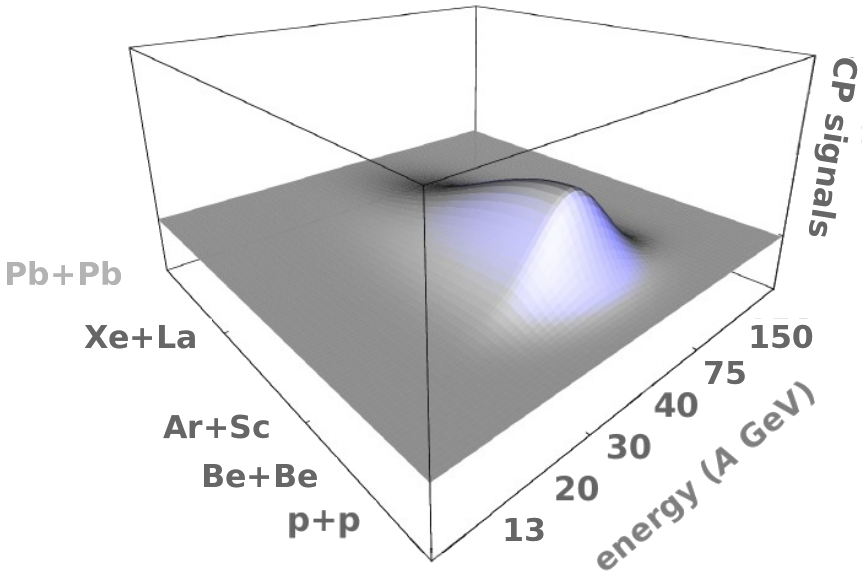}%
    \hspace*{3em}
    \includegraphics[align=c,height=0.18\textheight]{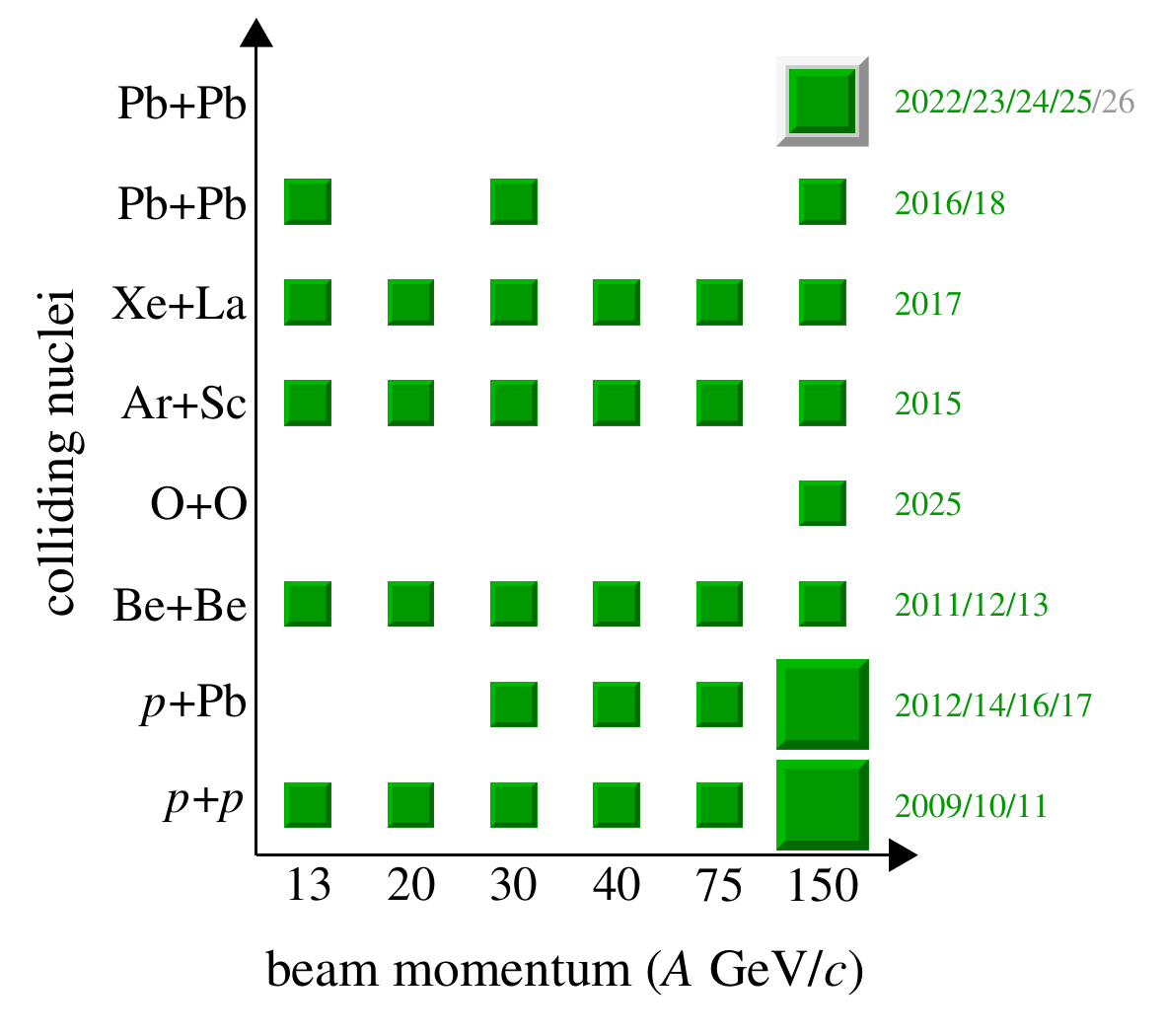}%
    \hspace*{3mm}%
 \end{center}
 \caption{(\emph{Left}) Schematic representation of ``hill of fluctuations'' expected around the CP; (\emph{Right}) Status of NA61/SHINE scan in collision energy and system size.}
 \label{fig:SHINE}
\end{figure}

NA61/SHINE \cite{NA61_experiment} is the second largest fixed-target experiment at CERN. As a direct, upgraded continuation of its predecessor NA49, the detector offers a nearly continuous experimental data flow for three decades. Its research program includes strong interaction physics as well as measurements for neutrino (J-PARC, Fermilab) and cosmic-ray studies (like the Pierre Auger Observatory, Telescope Array and IceTop). In this paper, only a selected set of most recent results related
to the NA61/SHINE critical point search program will be addressed.

A key question in the study of non-perturbative QCD is to determine the structure of the QCD phase diagram, that is, the location and type of phase transition boundaries between the various states of strongly interacting matter, as a function of temperature $T$ and baryochemical potential $\mu_B$ (or nuclear density, $n_B$). Combined evidence from lattice QCD and effective models indicates a smooth cross-over between hadronic matter and quark-gluon plasma at high $T$ and low $\mu_B$, and a 1\ts{st} order transition at low $T$ and high $\mu_B$, ending at a critical point (CP), in the
vicinity of which a 2\ts{nd} order transition is widely expected to occur.

At the vicinity of the CP, non-monotonic fluctuations of various observables~\cite{Gazdzicki-Seyboth:hill}, as well as scale-invariant (power-law) particle correlations~\cite{Antoniou:2001,Antoniou:2005,Antoniou:2006} are expected to occur (Fig.~\ref{fig:SHINE} (\emph{left})). Correlations in configuration space are projected to momentum space via quantum statistics and collective flow, where they can be observed by collision experiments. In order to do so, one must perform a carefully stepped scan in parameters that can be controlled experimentally, such as collision energy, nuclear mass number, centrality, and others. By varying such parameters, the freeze-out conditions of the colliding systems in $(T , \mu_B)$ vary accordingly~\cite{Becattini_Marek:freezout}. \shine has completed a two-dimensional scan in collision energy ($\sqrt{s_{NN}} = 5 - 17$~GeV) and system size ($p$+$p$, $p$+Pb, Be+Be, O+O, Ar+Sc, Xe+La, Pb+Pb), Fig.~\ref{fig:SHINE} (\emph{right}); it is therefore in a unique position to pursue the search for critical point signatures in nuclear collisions.

\section{Net-electric charge fluctuations}

Fluctuations of conserved
charges (electric, strangeness or baryon number) are of particular interest in the \shine energy-system size scan~\cite{StatusReport:2022}.

To compare fluctuations in systems of different sizes, one should use quantities insensitive
to system volume, i.e. intensive quantities. They are constructed by division of cumulants $\kappa_i$ of
the measured multiplicity distribution (up to fourth order), where $i$ is the order of the cumulant. In case of net-charge, cumulant ratios are defined to be $\kappa_2[h^+ - h^-] /$\break $ (\kappa_1[h^+] +  \kappa_1[h^-])$, $\kappa_3[h^+ - h^-]/\kappa_1[h^+ - h^-]$
and $\kappa_4[h^+ - h^-]/\kappa_2[h^+ - h^-]$, where $h^{+ (-)}$ stands for positively (negatively) charged hadrons.

\begin{figure}
 \begin{center}
 \includegraphics[width=.97\textwidth]{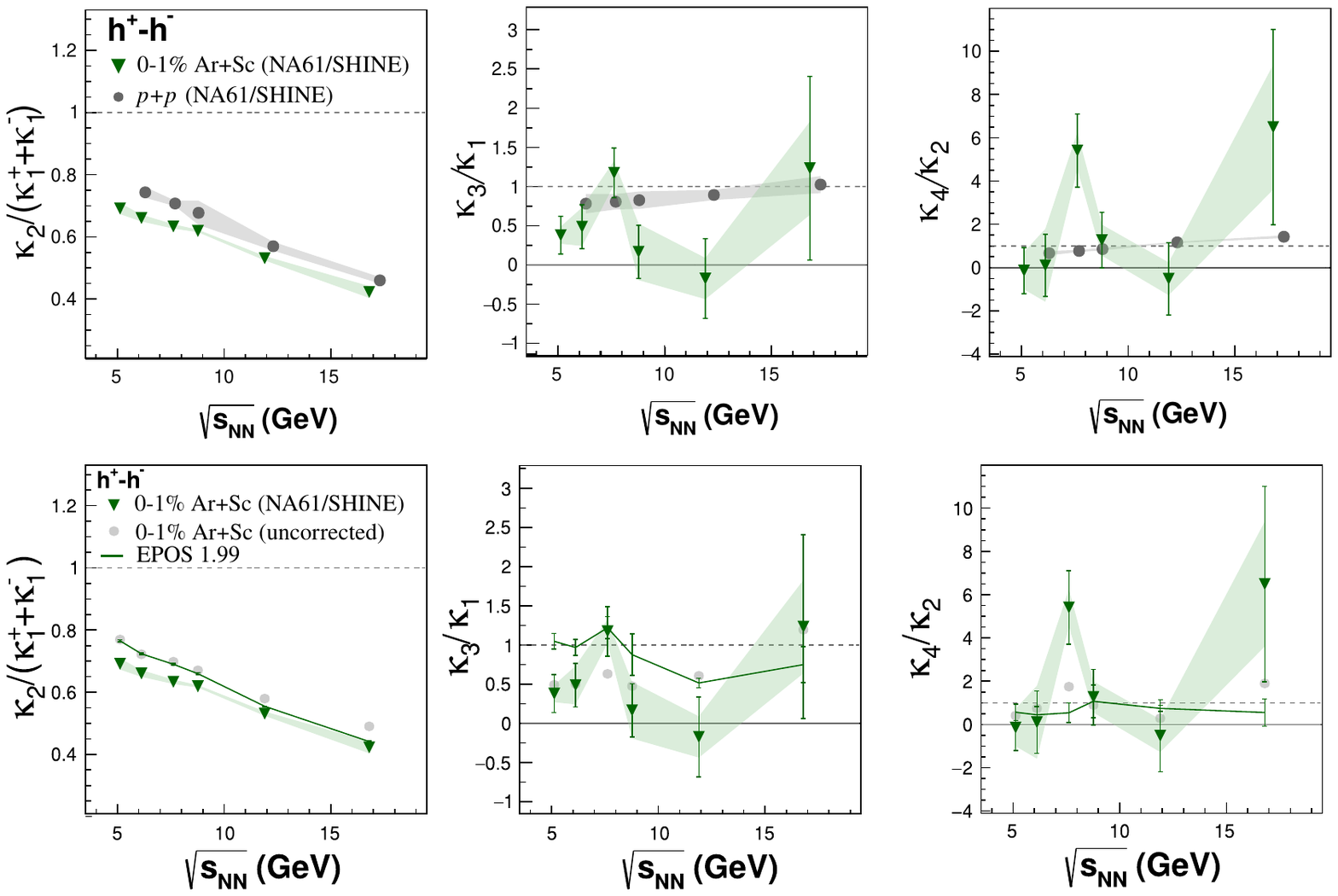}
 \end{center}
 \caption{(\emph{Top Row}) The energy dependence of
 $\kappa_2  / \left(\kappa_1^+ +  \kappa_1^- \right)$ (left),
 $\kappa_3 / \kappa_1$ (middle), and $\kappa_4 / \kappa_2$
 (right) of net-charge in the 1\% most central ${}^{40}$Ar+${}^{45}$Sc (triangles) and inelastic $p$+$p$ \shine interactions (gray circles); (\emph{Bottom Row}) The same ratios of cumulants as in Top Row for ${}^{40}$Ar+${}^{45}$Sc interactions (color triangles) compared with \texttt{\textsc{Epos1.99}} model
predictions (solid color lines). Color triangles correspond to quantities obtained from corrected distributions, while gray circles correspond to those obtained from uncorrected distributions (statistical uncertainties not indicated). The error bars correspond to statistical uncertainties via bootstrap, while the color bands correspond to systematic uncertainties. The dashed line at unity corresponds to the reference value of the Skellam distribution. The solid line at zero corresponds to the case with no fluctuations in the system. Figure adapted from Ref.~\cite{NA61SHINE:net_electric_ArSc}.}
 \label{fig:net_charge}
\end{figure}

Figure~\ref{fig:net_charge} shows the system size and energy dependence of second, third and fourth order cumulant ratio of net-electric charge in $p$+$p$ as well as central Ar$+$Sc interactions~\cite{NA61SHINE:net_electric_ArSc} (\emph{top row}), as well as its comparison with the \texttt{\textsc{Epos1.99}} model predictions (\emph{bottom row}). \texttt{\textsc{Epos1.99}} reproduces the magnitude of the signal observed in the data, with the largest deviations occurring at mid-SPS energies. There is a hint of non-monotonic behavior in $\kappa_3/\kappa_1$ and $\kappa_4/\kappa_2$ for central Ar+Sc at mid-SPS energies; however, statistical uncertainties are large. So far, no significant non-monotonic signal has been observed. More detailed studies are needed.

\section{Femtoscopy analysis}

The study of Bose-Einstein correlations (femtoscopy) in pairs of identical bosons (e.g. pions) produced in an ion collision can reveal the space-time structure of the hadron emitting source~\cite{Csorgo:2005it}. In particular,  for small to intermediate systems, we compare measurements with source calculations based on L\'evy-distributed sources, to
explore the pair transverse mass dependence of the L\'evy source parameters. The L\'evy exponent $\alpha$ is of particular significance, as it characterizes the shape of the source and is sensitive to the system freezing out in the vicinity of the CP.

\begin{figure}
 \begin{center}
    \includegraphics[align=c,height=0.13\textheight]{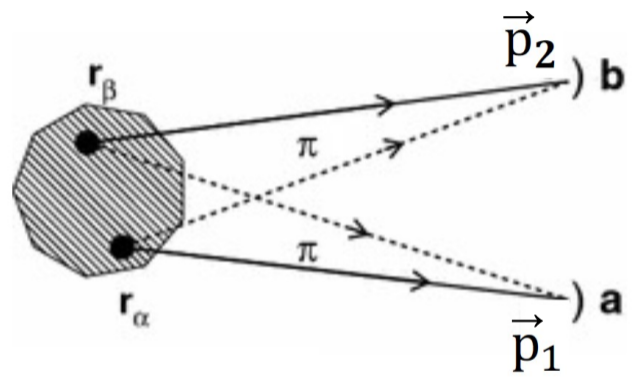}%
    \hspace*{3mm}%
    \includegraphics[align=c,height=0.25\textheight]{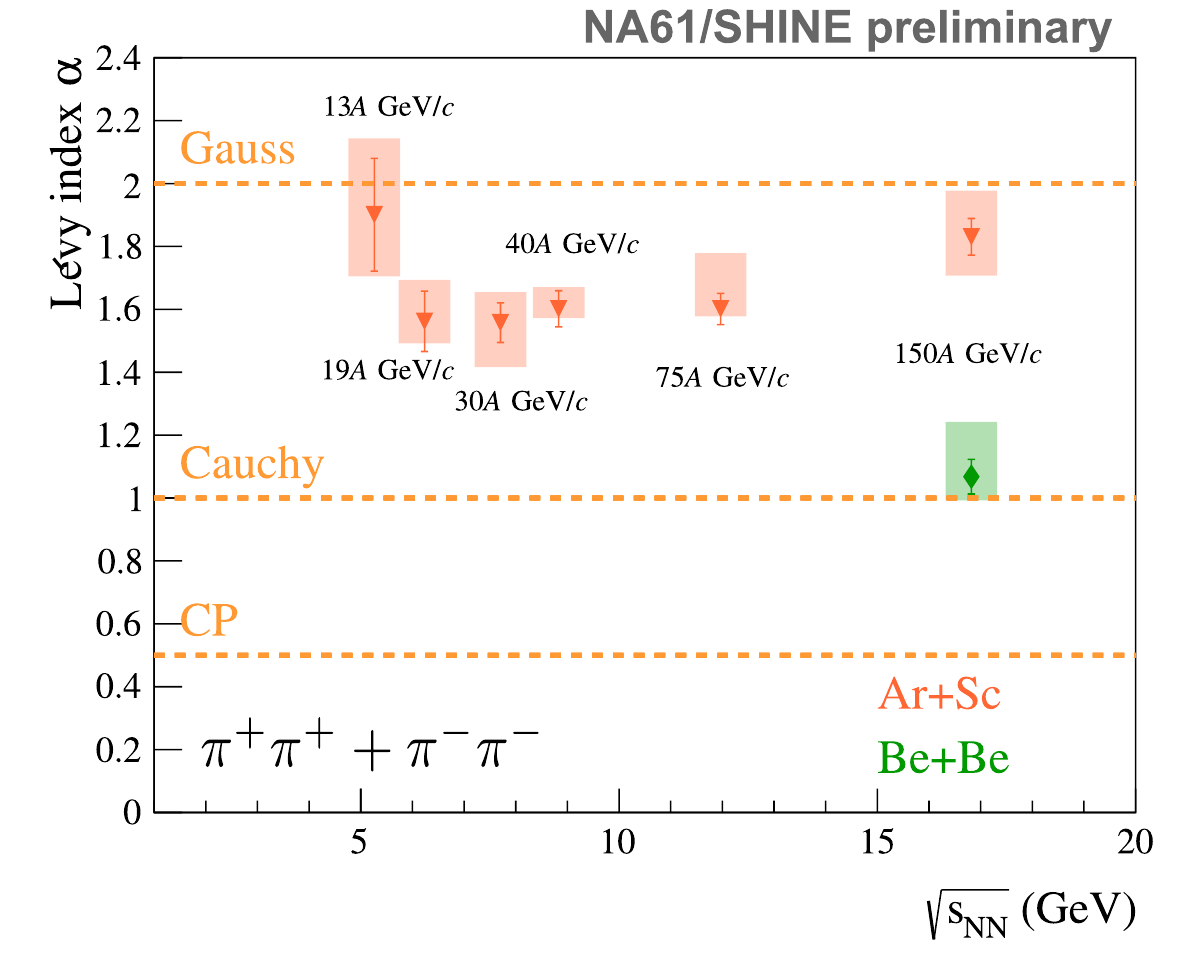}%
 \end{center}
 \caption{(\emph{Left}) Schematic representation of a correlated pair of pions emitted from a source; (\emph{Right}) Values of $\alpha$ obtained by fitting the pion pair correlation function, Eq.(\ref{eq:femtoscopic_correlation_function}), for pions produced in 0--20\% central Be$+$Be at 150$A$~GeV/$c$ and 0--10\% central Ar$+$Sc collisions at all
available energies, as a function of $\sqrt{s_{NN}}$. Special cases corresponding to a Gaussian $(\alpha = 2)$ or a Cauchy
$(\alpha = 1)$ source are shown, as well as $\alpha = 0.5$, which corresponds to the critical point~\cite{Csorgo:2005it}. Boxes denote systematic uncertainties, bars represent statistical uncertainties. Figure from Ref~\cite{Porfy:2024}.}
 \label{fig:femtoscopy}
\end{figure}

The femtoscopic correlation function fit to the data takes the form:

\begin{equation}
 C(q) = 1 + \lambda \cdot e^{-\left( q R \right)^\alpha},
 \label{eq:femtoscopic_correlation_function}
\end{equation}
where $q = |\mathbf{p}_1 - \mathbf{p}_2|$ is the momentum difference between bosons in the pair (Fig.~\ref{fig:femtoscopy} (\emph{left})), $R$ is the L\'evy scale parameter, $\lambda$ is the correlation strength, and $\alpha$ is the L\'evy exponent (index) which is a shape parameter of the distribution (for $\alpha = 2$, we get a Gaussian distribution).

Figure~\ref{fig:femtoscopy} (\emph{right}) shows the L\'evy index $\alpha$ values fitted to $\pi^+\pi^+$ and $\pi^-\pi^-$ pairs produced at different SPS collision energies, for Be$+$Be and Ar$+$Sc collisions observed by the \shine experiment. The recent Ar$+$Sc results are close to Gaussian, and far from the CP predicted value. No indication of the critical point is observed at any of the investigated energies.

\section{Intermittency analysis}

A characteristic feature of a second-order phase transition is the divergence of the correlation length of the physical system, leading to a scale-invariant system with long-range correlations.  The density-density correlation function of the phase transition order parameter (for example, the chiral $\sigma$-condensate or the net baryon density $n_B$ \cite{Hatta-Stephanov:2003}) assumes a power-law form, with critical exponents determined by the 3D Ising universality class~\cite{Antoniou:2001,Antoniou:2005,Antoniou:2006}. Observables tailored for probing such power-law behavior are the scaled factorial moments (SFMs) $F_r(M)$ in transverse momentum space,

\begin{multline}
  F_r(M) \equiv\\ \left\langle \frac{1}{M^2} \displaystyle\sum_{m=1}^{M^2} n_m (n_m-1)\ldots (n_m - r + 1) \right\rangle \Bigg{/} \left\langle\frac{1}{M^2} \displaystyle\sum_{m=1}^{M^2} n_m \right\rangle^r,
 \label{eq:fact_moments_r}
\end{multline}
where $n_m$ is the number of particles in a transverse momentum $(p_x ,\, p_y)$ bin in an event, $M^2$ is the total number of 2D bins, and $\langle \ldots \rangle$ denotes averaging over events. Intermittency is defined as a power-law scaling of $F_r(M)$ with the number of bins, or equivalently bin size: $F_r(M) \sim \left( M^2 \right)^{\phi_r}$, with $\phi_r$ being the $r$-th order intermittency index~\cite{Bialas:1986, Antoniou:2006}. For protons\footnote{Protons substitute for net baryons~\cite{Hatta-Stephanov:2003}, an approximation that is valid when antiprotons are much fewer than protons in events.} and $r=2$, $\phi_2 = 5/6$ is expected~\cite{Antoniou:2006}. When probing intermittent behavior, baseline correlations have to be subtracted; this can be done either through the cumulative variable transformation, imposing a uniform 1-particle $(p_x ,\, p_y)$ distribution~\cite{Bialas-Marek, NA61SHINE:Tobiasz_intermittency, NA61SHINE:2024ffp}, or by subtracting mixed event moments from those of data~\cite{NA49intermittency:2015}; the resulting subtracted scaled factorial moments carrying the possible critical signal are denoted by $\Delta F_r(M)$, or $\Delta F_r(M)_c$ for cumulative.

\begin{figure}
 \begin{center}
  \includegraphics[align=c,height=0.19\textheight]{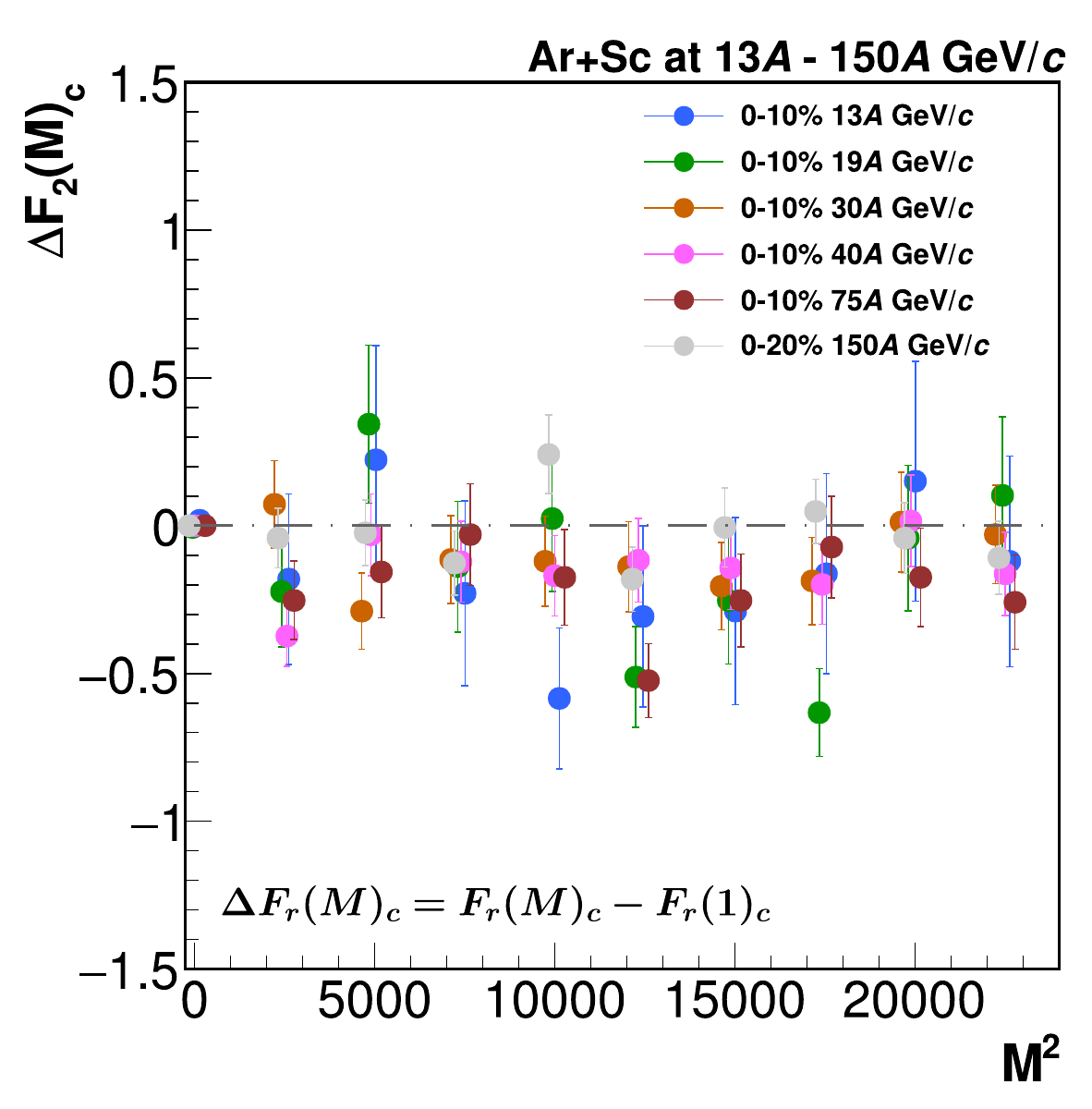}%
  \hspace*{2mm}%
  \includegraphics[align=c,height=0.19\textheight]{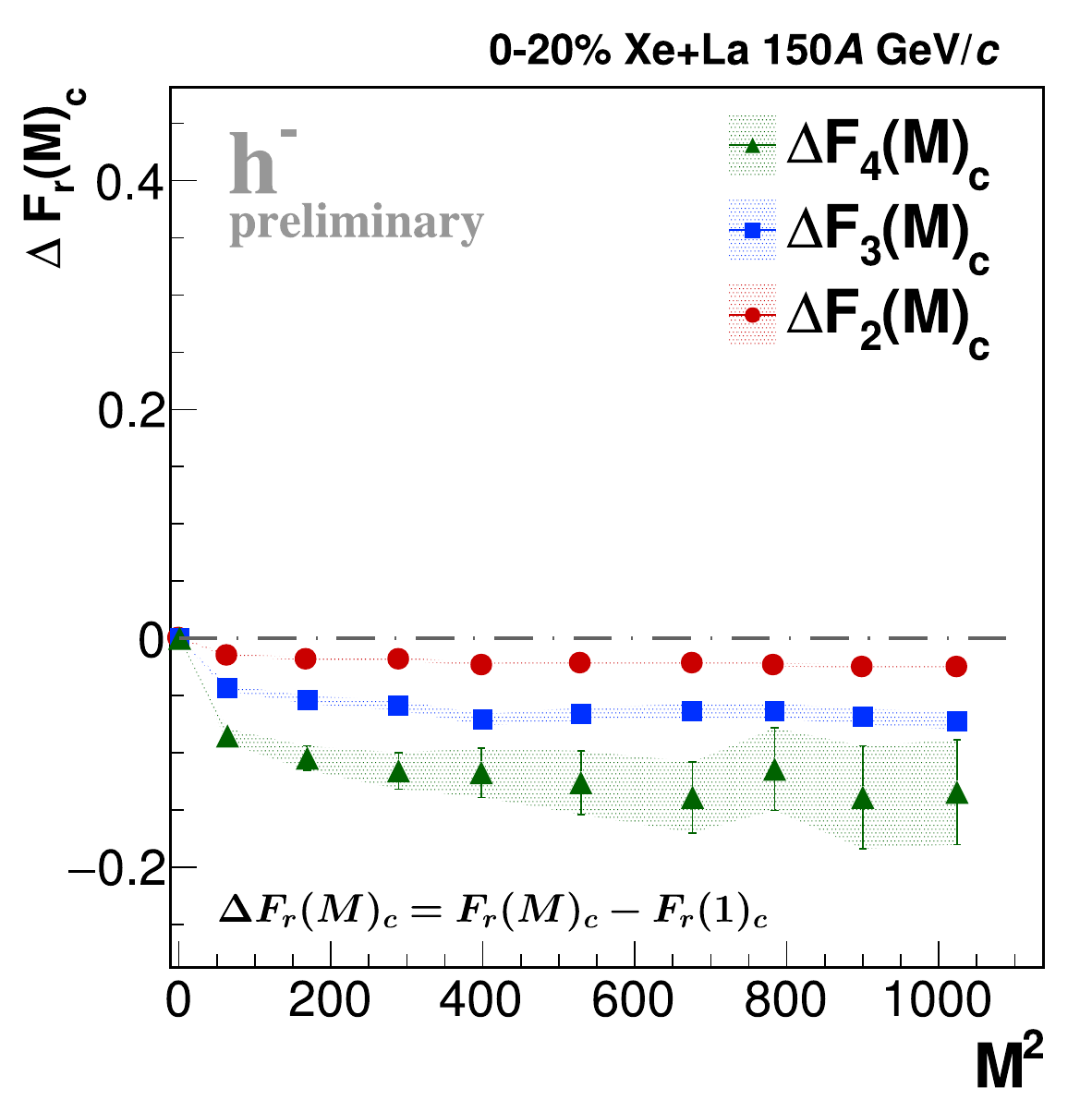}%
  \hspace*{2mm}%
  \includegraphics[align=c,height=0.19\textheight, trim={2mm 23mm 28mm 12mm},clip]{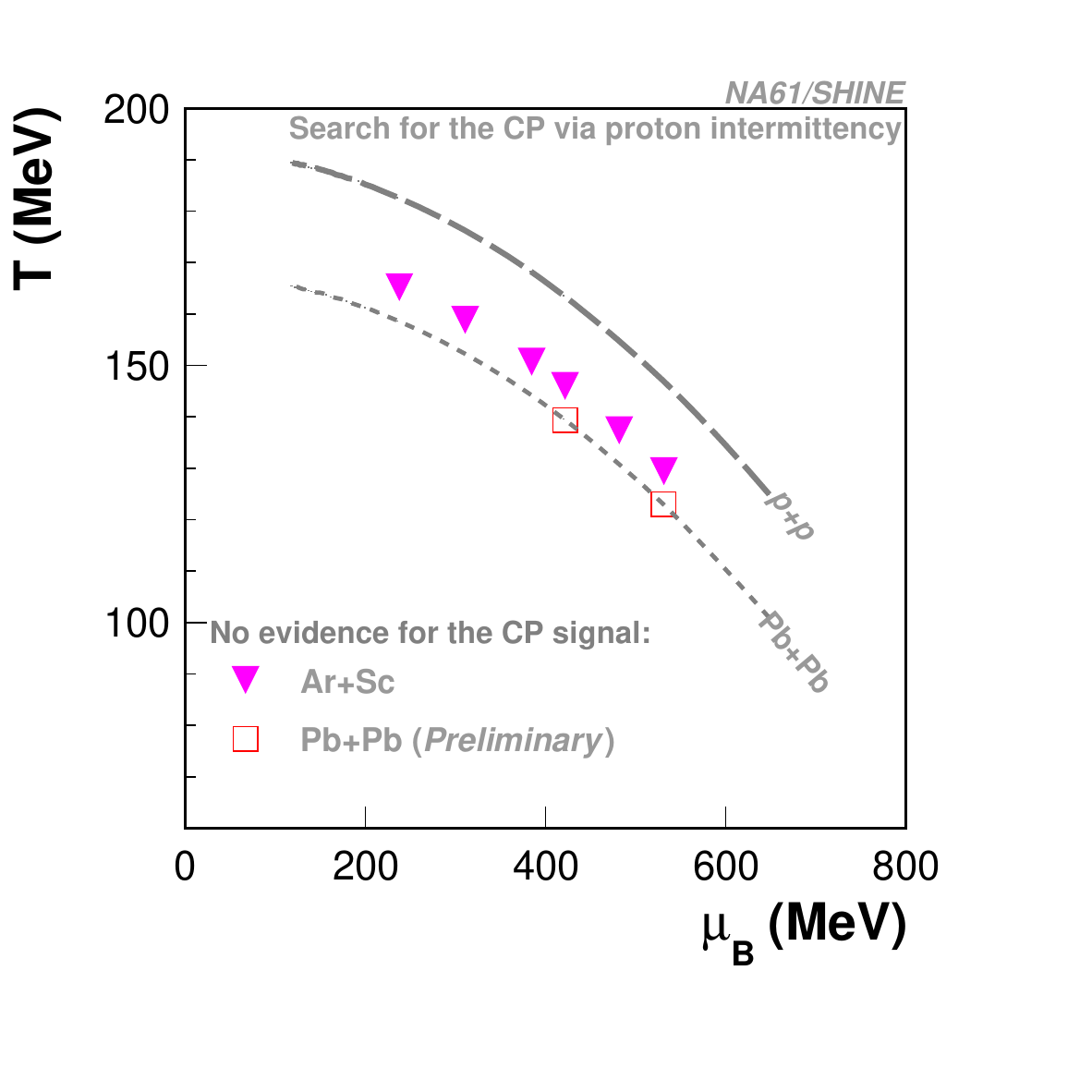}%
 \end{center}
 \caption{(\emph{Left}) Results on the proton multiplicity 2nd-order SFMs $\Delta F_2(M)_c$ for \shine Ar$+$Sc collisions in the beam momentum range $13A-150A$~GeV/$c$ ($\sqrt{s_{NN}} = 5 - 17$~GeV)~\cite{NA61SHINE:Tobiasz_intermittency,NA61SHINE:2024ffp}; (\emph{Middle}) 2nd through 4th-order SFMs for \shine negatively charged hadrons in Xe$+$La collisions at 150$A$~GeV/$c$ ($\sqrt{s_{NN}} = 16.8$~GeV)~\cite{Ortiz:2024mcg}; cumulative variables~\cite{Bialas-Marek} and independent bins are used; (\emph{Right}) Status of \shine CP search via proton intermittency~\cite{NA61SHINE:2024ffp}. Points indicate freeze-out conditions of respective reactions in $(T, \mu_B)$.
 }
 \label{fig:F2_cumulative}
\end{figure}

Figure~\ref{fig:F2_cumulative}~(\emph{left}) shows the behavior of proton multiplicity moments\break $\Delta F_2(M)_c$ for a number of \shine Ar$+$Sc collision datasets in the beam momentum range $13A-150A$~GeV/$c$ ($\sqrt{s_{NN}} = 5 - 17$~GeV). The number of transverse momentum space subdivisions (bins) ranges in $1^2 \leq M^2 \leq 150^2$, in a window $-1.5 \leq p_{x,y} \leq 1.5$~GeV/$c$ at mid-rapidity. No intermittency signal indicative of the CP is observed.

\shine also looked into negatively charged hadrons ($h^-$) intermittency in Xe$+$La collisions at 150$A$~GeV/$c$ ($\sqrt{s_{NN}} = 16.8$~GeV), as a check of the published STAR Collaboration intermittency results for all charged hadrons in Au$+$Au collisions~\cite{STAR:2023jpm}. Figure~\ref{fig:F2_cumulative}~(\emph{middle}) shows \shine $h^-$ results for $r=2-4$ SFMs. When a cumulative transform is applied to the momenta in order to remove baseline/spurious effects, no intermittency effect remains. The effect similarly vanishes when a short-range correlation cut is applied, suggesting HBT-like correlations are at the origin of the observed effect~\cite{Ortiz:2024mcg}.

Figure~\ref{fig:F2_cumulative}~(\emph{right}) summarizes \shine system size \& energy proton intermittency scan, showing the estimated freeze-out conditions for the analyzed systems; so far, no significant intermittency signal is observed.

\begin{figure}
 \begin{center}
  \includegraphics[align=c,height=0.17\textheight]{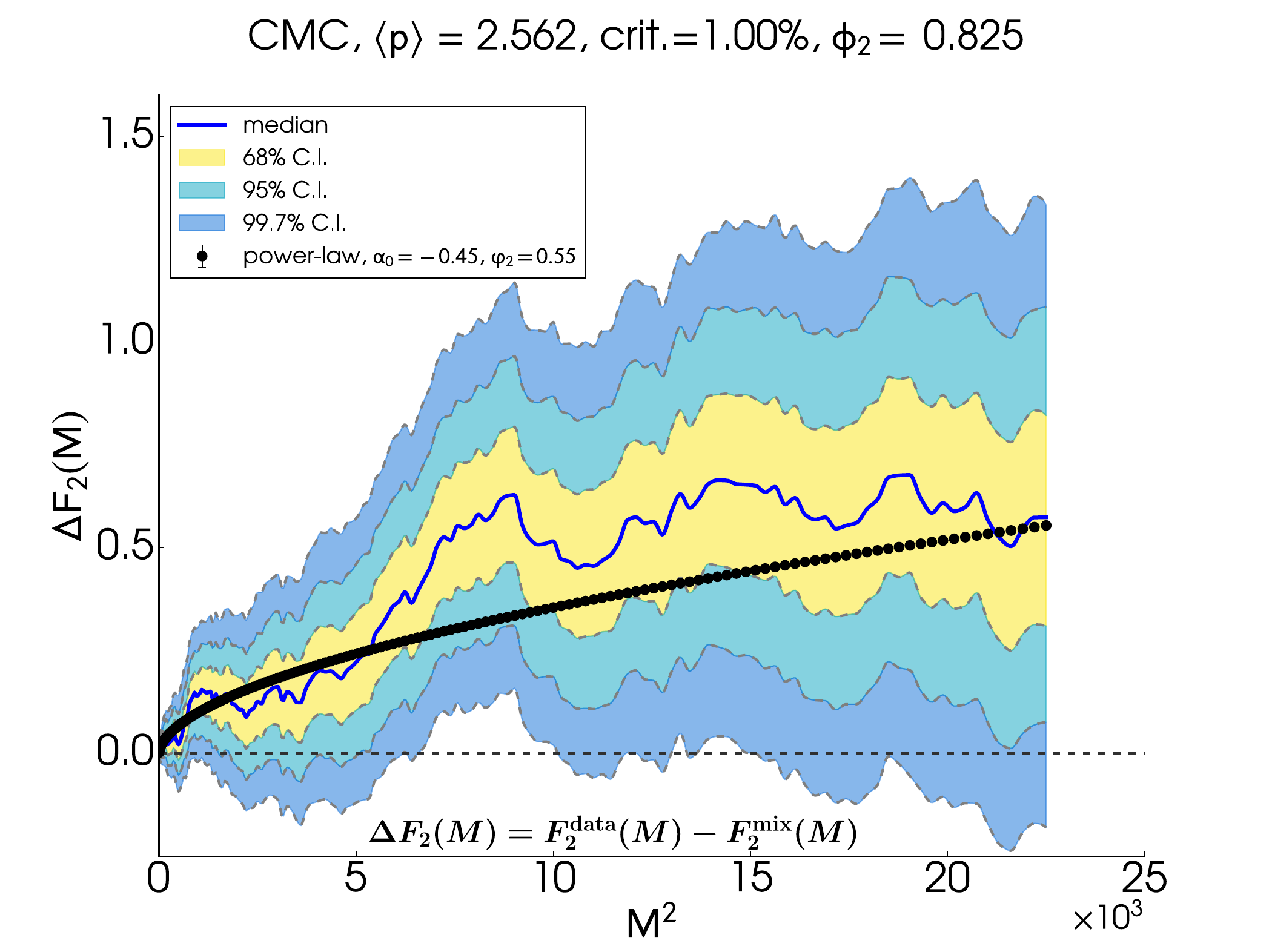}%
  \includegraphics[align=c,height=0.17\textheight]{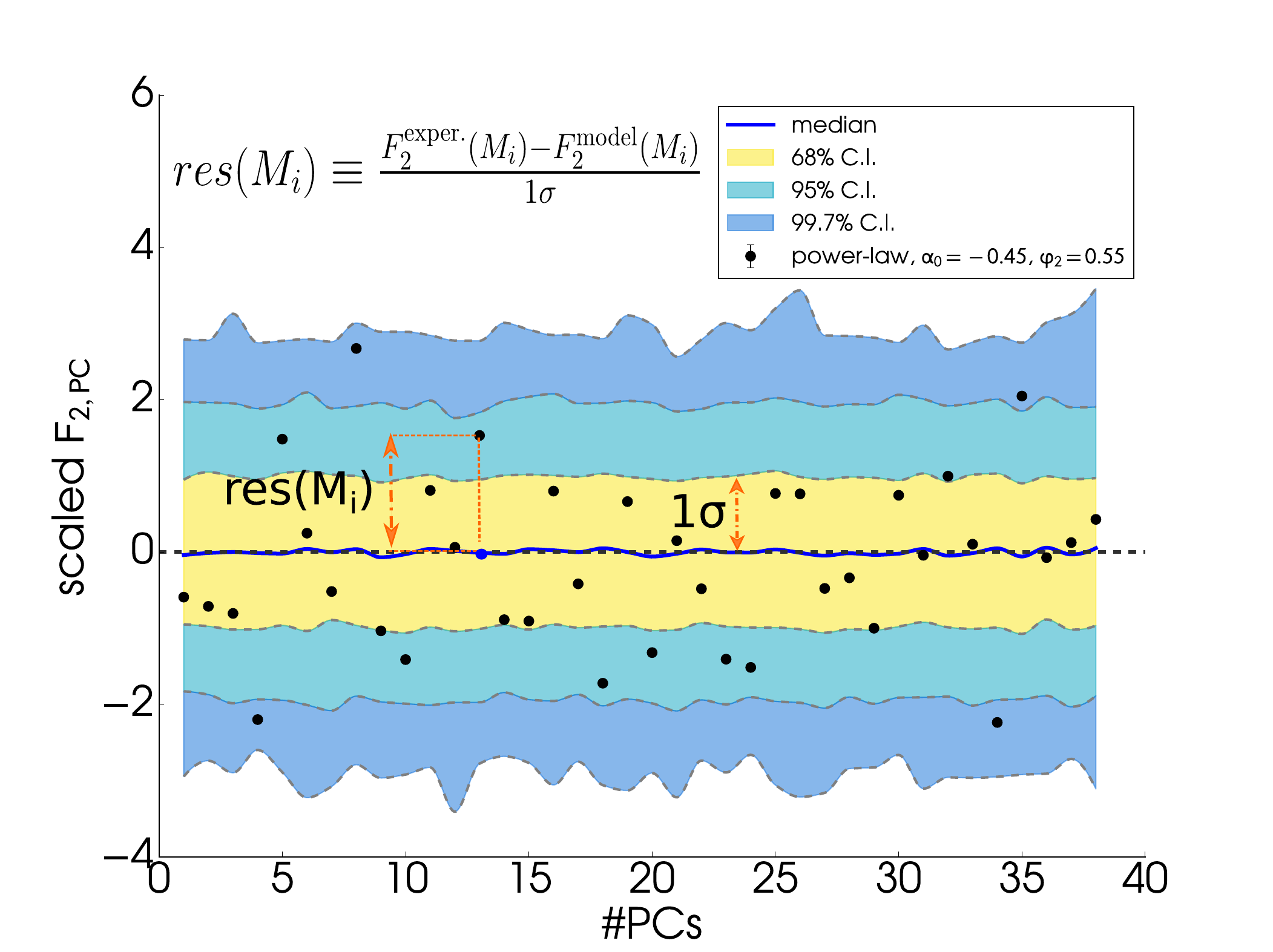}%
  \includegraphics[align=c,height=0.17\textheight]{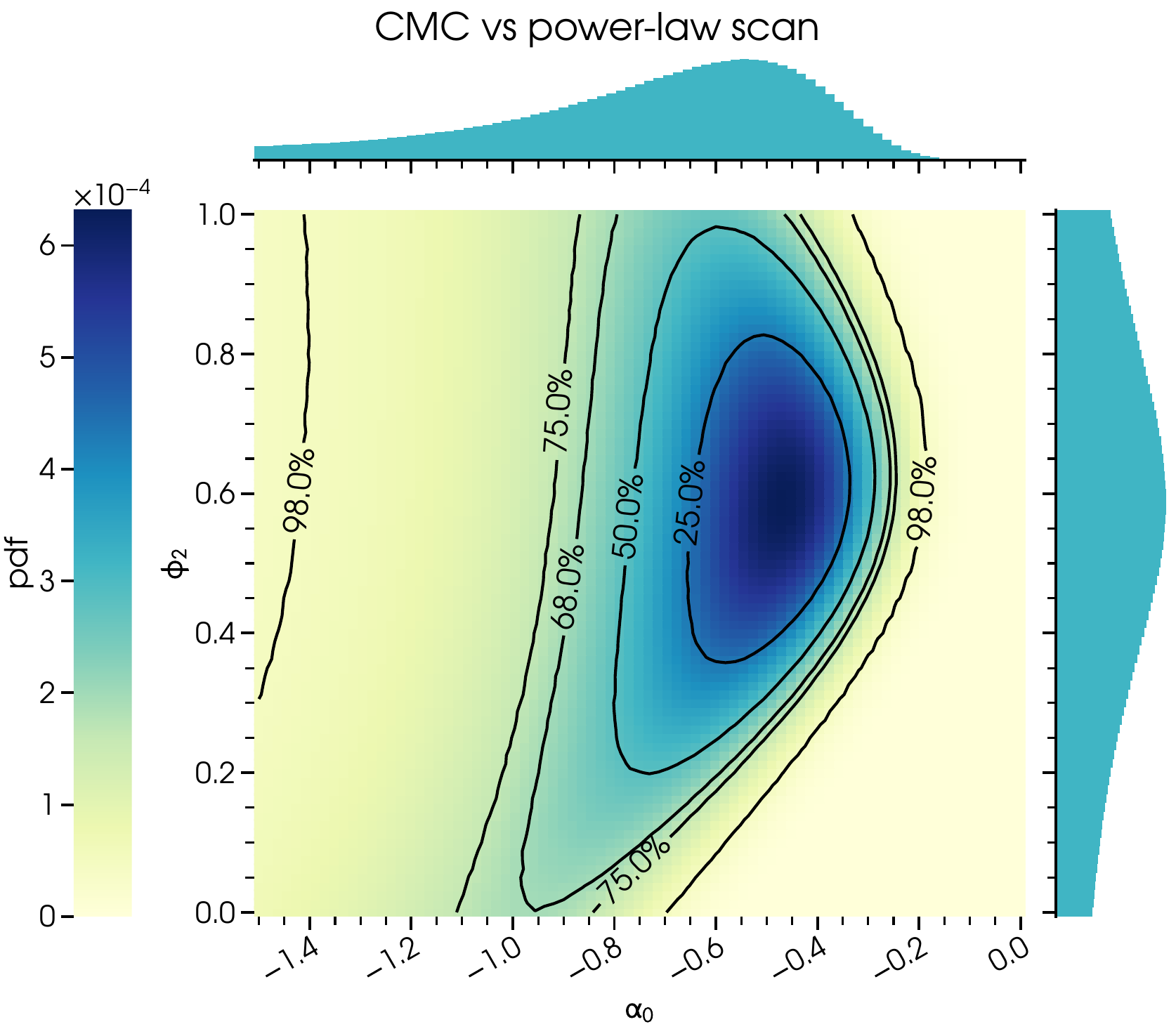}%
 \end{center}
 \caption{(\emph{Left}) Correlated proton $\Delta F_2(M)$ for $\sim 8$K bootstrap samples~\cite{Metzger} of a synthetic CMC data set (colored bands) corresponding to $\sim 400$K Ar$+$Sc-like collisions with 1\% critical protons. Black points represent a power-law model fit, with parameters $(\alpha_0, \phi_2)$; (\emph{Middle}) $\Delta F_2(M)$ for the same CMC data set, transformed to the PC coordinates; (\emph{Right}) Normalized likelihood distribution for a grid of parametrized power-law models against the same CMC data set. Figures adapted from Ref.~\cite{Davis:PCA}.}
 \label{fig:F2_correlated}
\end{figure}

The above presented \shine intermittency results utilize independent bins in order to avoid $M$-bin correlations, which invalidate fitting and comparison with models. This method has the drawback of reducing event statistics per bin, thus significantly inflating uncertainties. An alternative approach has been developed to handle bin correlations~\cite{Davis:PCA}, based on the statistical tool of Principal Component Analysis (PCA). Figure~\ref{fig:F2_correlated} illustrates how PCA works: a set of original moments generated via the Critical Monte Carlo (CMC) simulation~\cite{Antoniou:2006, Davis:PCA} (\emph{left}) are transformed into statistically independent linear combinations of the original $M$-bins, called the principal components (PCs) (\emph{middle}). A power-law candidate model (black points), ${\Delta}F_2^{\text{model}}(M)  \equiv  10^{\alpha_0} \left( \frac{M^2}{10^4} \right)^{\phi_2}$ is transformed in the same way, and its residuals are compared to the data bootstrap variance in the PCs. The result is a likelihood value for the model, which can be collected into a likelihood distribution of model parameters (\emph{right}). This allows the reliable extraction of $\phi_2$ and other parameter confidence intervals from correlated data, making use of the full event statistics. We expect this method to permit more reliable studies of intermittency phenomena, in \shine as well as other high-energy physics experiments.

\section{Summary}

In this work, we have provided an overview of the \shine critical point search program, including net-charge fluctuations, femtoscopy analysis, and intermittency analyses  on both protons and negatively charged hadrons, for a variety of system sizes and SPS collision energies.
Findings from \shine give no clear indication of the critical point so far; however, they point at the crucial importance of a proper understanding of non-critical (baseline) effects in the CP search.

We report a new development in intermittency analysis, which effectively solves the long-standing bin-by-bin correlation problem~\cite{Davis:PCA}; this allows the use of full statistics with correlated bins and reliable model scans to locate the best-fitting models to the data. Event statistics plays a crucial role, as intermittency analysis is particularly data-hungry.  Nevertheless, it is within the capabilities of the upgraded \shine detector to provide sufficient data for such analyses.

\begin{small}
\textbf{Acknowledgements:} The present work was supported by the Polish Ministry of Science and Higher Education (2025/WK/05).
\end{small}

\bibliographystyle{wocl}
\bibliography{my-bib-database}

\end{document}